\AtBeginDocument{%
  }
    
\documentclass[sigconf]{acmart}
\usepackage{algorithmic}
\usepackage{graphicx}
\usepackage{textcomp}
\usepackage{xcolor}
\usepackage{hyperref}
\usepackage{soul}
\usepackage{physics,array}
\usepackage{multirow}
\usepackage{tikz}

\setcopyright{acmlicensed}
\copyrightyear{2018}
\acmYear{2018}
\acmDOI{XXXXXXX.XXXXXXX}

\renewcommand\footnotetextcopyrightpermission[1]{}

\acmBooktitle{Woodstock '18: ACM Symposium on Neural Gaze Detection,
 June 03--05, 2018, Woodstock, NY}
\acmISBN{978-1-4503-XXXX-X/18/06}

\begin{document}
\fancyhead{}

\title{Q-Embroidery: A Study on Weaving Quantum Error Correction into the Fabric of Quantum Classifiers}


\author{Avimita Chatterjee}
\email{amc8313@psu.edu}
\affiliation{%
  \institution{The Pennsylvania State University}
  \city{State College}
  \state{Pennsylvania}
  \country{USA}
  \postcode{16801}
}

\author{Debarshi Kundu}
\email{dqk5620@psu.edu}
\affiliation{%
  \institution{The Pennsylvania State University}
  \city{State College}
  \state{Pennsylvania}
  \country{USA}
  \postcode{16801}
}

\author{Swaroop Ghosh}
\email{szg212@psu.edu}
\affiliation{%
  \institution{The Pennsylvania State University}
  \city{State College}
  \state{Pennsylvania}
  \country{USA}
}

\renewcommand{\shortauthors}{Anonymous et al.}

\begin{abstract}
  Quantum computing holds transformative potential for various fields, yet its practical application is hindered by the susceptibility to errors. This study makes a pioneering contribution by applying quantum error correction codes (QECCs) for complex, multi-qubit classification tasks. We implement 1-qubit and 2-qubit quantum classifiers with QECCs, specifically the Steane code, and the distance 3 \& 5 surface codes to analyze 2-dimensional and 4-dimensional datasets. This research uniquely evaluates the performance of these QECCs in enhancing the robustness and accuracy of quantum classifiers against various physical errors, including bit-flip, phase-flip, and depolarizing errors. The results emphasize that the effectiveness of a QECC in practical scenarios depends on various factors, including qubit availability, desired accuracy, and the specific types and levels of physical errors, rather than solely on theoretical superiority.
\end{abstract}




\maketitle

\section{Introduction} \label{sec:intro}

Quantum computing represents a significant leap forward in computational capabilities, offering the potential to solve complex problems that are intractable for classical computers. By leveraging the principles of quantum mechanics, such as superposition, entanglement and interference, quantum computers can perform certain calculations much more efficiently than their classical counterparts. This makes them particularly useful for tasks like cryptography, material science simulations, and optimization problems, where they can potentially provide solutions exponentially faster \cite{montanaro2016quantum, nielsen2010quantum}.

Quantum machine learning (QML) merges quantum computing with machine learning, potentially speeding up data processing and analysis tasks. This synergy could revolutionize artificial intelligence by improving big data handling, complex computations, and the creation of innovative learning algorithms. QML promises a future where quantum algorithms surpass classical methods in many areas \cite{schuld2015introduction, li2022recent}.

As quantum computing advances, the importance of quantum error correction becomes paramount. Qubits are highly susceptible to errors due to quantum noise, which can significantly undermine the reliability of quantum computations. Quantum error correction schemes are crucial for protecting information stored in qubits from errors, thereby ensuring the practical usability and scalability of quantum computers. These error correction methods enable the construction of fault-tolerant quantum computers, making them more resilient to errors and viable for real-world applications \cite{terhal2015quantum}.


\subsection{Motivation}

The advancement of quantum computing has brought about a range of quantum error correction codes, primarily aimed at protecting operations on single qubits from errors. Yet, there is a marked deficiency in research concerning their application to broader quantum circuits and more complex practical quantum operations. This shortfall is predominantly due to the complexities involved in extending these codes beyond simple qubit protection. Notably, the application of quantum error correction codes in real-world quantum tasks, such as quantum classifiers, remains largely untapped. This gap signals a pivotal opportunity for research to evaluate how these codes can boost both the performance and reliability of quantum circuits in executing real-world tasks, thereby making a compelling case for their broader application and potential impact.

\subsection{Contribution}

This research, to the best of our knowledge, is the first study to systematically apply QECC to quantum classifiers. We explore the application of three distinct QECCs: the Steane Code \cite{steane1996multiple}, and two variations of the surface code with distances $3$ and $5$ \cite{krinner2022realizing}. Our investigation spans a spectrum of classifier complexities, ranging from a simpler 1-qubit classifier applied to a synthetic 2D dataset to a more intricate 2-qubit classifier designed for a synthetic 4D dataset. A key element of our research is the detailed comparison of quantum circuits' performance pre- and post-QECC application and the computational overhead each code introduces. Additionally, we conduct experiments across three error modes: depolarizing, bit and phase flips, and a combination thereof, with varying intensities of physical noise to empirically show QECC's substantial improvement on both 1-qubit and 2-qubit classifier performance.

The importance of this seemingly simple task cannot be overstated. Implementing QECCs in even a single qubit significantly increases the need for additional qubits and gates, leading to exponential growth in overhead as circuit complexity rises. This challenge is further compounded by the need for theoretical transversal gates, such as lattice surgery, for multi-qubit gates. By concentrating on a straightforward classifier, we aim to contain this overhead and complexity, proving that success with basic models today can facilitate future progress with more complex systems as QECC technology evolves. 
Our findings emphasize the importance of carefully choosing QECCs based on the specific needs and constraints of quantum tasks, beyond just theoretical preferences. We assume readers have a basic understanding of QECCs, especially the Steane and Surface codes' mechanics and applications as discussed in the existing literature \cite{chatterjee2024magic, steane1996multiple, krinner2022realizing}. 

\subsection{Paper Structure}    

The paper starts with a background on quantum error correction and classifiers (Section \ref{sec:background}), outlines our methodology including dataset and quantum classifier details, error models, QECCs, and evaluation metrics (Section \ref{sec:methodology}), analyzes the impact of physical errors and QECC effectiveness (Section \ref{sec:exp_result}), discusses limitations and challenges (Section \ref{sec:lim}), and concludes with key findings (Section \ref{sec:conclusion}).
\section{Background} \label{sec:background}

\subsection{Quantum Error Correction}

Quantum Error-Correcting Codes (QECCs) stand in stark contrast to classical error correction techniques, as they safeguard information encoded in quantum states, which are defined by unique characteristics such as superposition and entanglement \cite{nielsen2010quantum}. The delicate nature of quantum systems means that quantum states are highly susceptible to disruption from external noise \cite{preskill1998reliable}, a vulnerability that poses a major challenge for consistent quantum computation and data storage. QECCs address this challenge effectively \cite{gottesman1997stabilizer}.

By spreading quantum information across multiple qubits, QECCs enable error detection and correction without necessitating a direct measurement of the quantum state, thus remaining in compliance with the no-cloning theorem \cite{wootters1982single}. Quantum computing faces primarily bit-flip and phase-flip errors, alongside more intricate errors that combine these two \cite{shor1995scheme}. Through leveraging entangled states and collective measurements, QECCs are adept at correcting such errors, enhancing the robustness and reliability of quantum information processing \cite{calderbank1996good}.

\subsection{Quantum Classifiers}

Quantum classifiers utilize quantum computing principles like superposition, entanglement, and interference to offer advanced data processing and analysis capabilities for machine learning tasks. These quantum properties enable the simultaneous representation and processing of extensive data combinations, intricate data correlation capture, and enhanced classification accuracy. Classifiers are mainly divided into Variational Quantum Classifiers (VQCs), which employ parameterized quantum circuits optimized through classical feedback loops, and Quantum Kernel Methods, which project input data into a high-dimensional quantum space for analysis \cite{schuld2015introduction, li2022recent}.

The potential applications of quantum classifiers span various fields, including drug discovery for precise molecular structure analysis, finance for portfolio optimization and fraud detection, and cybersecurity for identifying complex threats beyond classical computing's reach. Additionally, they could improve logistics and supply chain management by more efficiently solving complex optimization problems. As quantum technology progresses, its integration could lead to breakthroughs in these areas by offering solutions currently beyond classical computational methods' capabilities \cite{biamonte2017quantum}.
\section{Methodology} \label{sec:methodology}

\subsection{Dataset Description}


Recognizing challenges like exponential qubit and gate increases in QECC-enhanced circuits and the need for advanced transversal gates for multi-qubit circuits, we have limited our scope to one- and two-qubit quantum classifiers, thus, developing tailored synthetic datasets to suit their capabilities.

\begin{figure}
    \centering
    \includegraphics[width=1\linewidth]{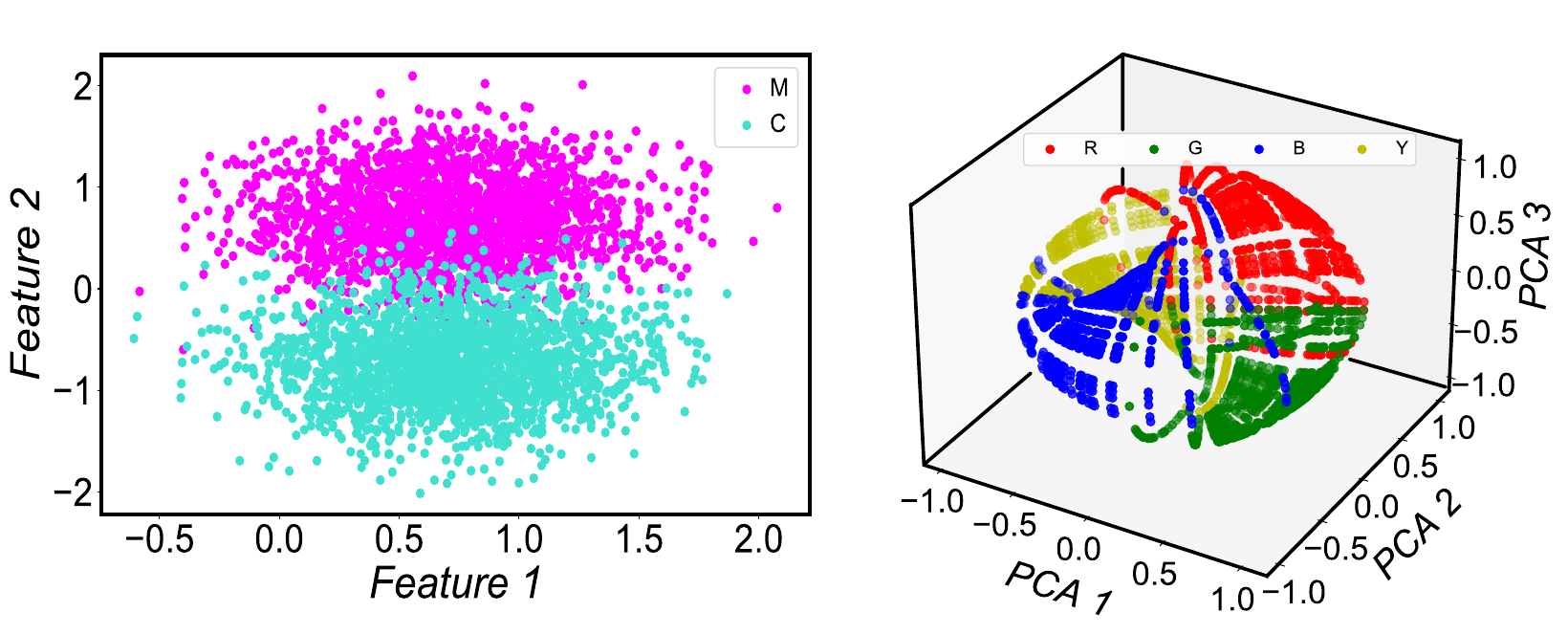}
    \vspace{-20pt}
    \caption{\textbf{Classical datasets for quantum classification:} 
    The left figure shows a two-dimensional dataset in magenta and cyan for the one-qubit classifier. The right figure presents a four-dimensional dataset in red, blue, green, and yellow for the two-qubit classifier, simplified to three dimensions using Principal Component Analysis (PCA) for visualization.
    }
    \label{fig:dataset}
\end{figure}

The initial dataset is fairly straightforward, containing $2048$ two-dimensional data points divided equally into two color classes: magenta and cyan. This dataset is specifically designed for the one-qubit classifier and is depicted in Fig. \ref{fig:dataset} (left). It provides a foundational platform for testing and refining the classifier's capabilities. In contrast, our second dataset is more elaborate, with $4096$ four-dimensional data points distributed equally among four color classes: red, blue, green, and yellow. To overcome challenges in visualization due to complexity, we applied Principal Component Analysis (PCA) to reduce the dataset's dimensions, preserving key variations. This approach allowed us to simplify the data into a three-dimensional format for easier visualization and interpretation, as shown in Fig. \ref{fig:dataset} (right).

An essential part of preparing these datasets for the quantum classifiers involves normalizing the data points into a quantum state format. This conversion is critical for ensuring that classical data can be effectively processed by quantum classifiers, allowing for efficient training, testing, and validation of the classifiers.

\subsection{Quantum Classifier Implementation}

\begin{figure}
    \centering
    \includegraphics[width=1\linewidth]{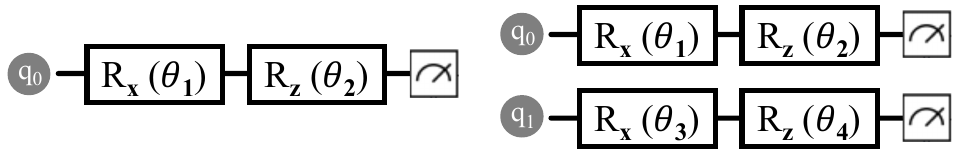}
    \vspace{-20pt}
    \caption{\textbf{Overview of quantum classifier circuits:} 
    The left figure shows a one-qubit classifier circuit for a two-dimensional dataset (Fig. \ref{fig:dataset} left), while the right figure depicts a two-qubit classifier circuit for a four-dimensional dataset (Fig. \ref{fig:dataset} right).
    }
    \label{fig:class_circuit}
    \vspace{-10pt}
\end{figure}

The one-qubit classifier operates using two rotational gates, one along the $X$ axis and the other along the $Z$ axis, requiring optimization of two parameters corresponding to these rotations. Conversely, the two-qubit classifier extends this architecture to accommodate two qubits, incorporating four rotational gates and necessitating the training of four parameters. The circuit designs for both classifiers are illustrated in Fig. \ref{fig:class_circuit}.

We partitioned the original dataset, allocating $80\%$ for training and the remaining $20\%$ for testing, employing k-fold validation to enhance the robustness of our findings. The performance metrics presented in the following sections are based on testing accuracy. 
The integration of rotational gates in error correction algorithms and circuits is challenging due to their sensitivity to quantum noise and the need for precise parameter control essential for quantum operation fidelity \cite{jayashankar2023quantum}. This necessitates transforming parameterized circuits, dependent on rotational gate fine-tuning, into non-parameterized versions. For this transformation, we employ the `Greedy-PQC-Optimization' method \cite{phalak2024non}, which is designed to systematically synthesize parameterized quantum circuits into their non-parameterized equivalents, thereby sidestepping the complexities associated with parameter tuning. The comparison between the original parameterized circuits and their synthesized non-parameterized forms, focusing on any variances in accuracy and the extent of accuracy reduction following the conversion is shown in table \ref{tab:all_basic_class}. Notably, we find that the decrease in accuracy is minimal, underscoring the effectiveness of the `Greedy-PQC-Optimization' process in maintaining computational performance while enhancing error resilience. Consequently, all further experiments, data analyses, and accuracy reports in this study will be based on these synthesized, non-parameterized circuits.

\begin{table}[]
\centering
\caption{Comparison of original and synthesized quantum classifier circuits.}
\begin{tabular}{cc||cc||cc}
\multicolumn{2}{c||}{\textbf{Original}}                                        & \multicolumn{1}{c|}{\textbf{Qubit}} & \textbf{Gate} & \multicolumn{1}{c|}{\textbf{Acc \%}} & \textbf{Reduce}     \\ \hline \hline
\multicolumn{1}{c|}{\multirow{2}{*}{\textbf{1-qubit}}} & \textbf{Original}    & 1                                   & 2             & 92.521                               & \multirow{2}{*}{1.285 \%} \\
\multicolumn{1}{c|}{}                                  & \textbf{Synthesized} & 1                                   & 1             & 91.328                               &                        \\ \hline
\multicolumn{1}{c|}{\multirow{2}{*}{\textbf{2-qubit}}} & \textbf{Original}    & 2                                   & 4             & 86.219                               & \multirow{2}{*}{2.235 \%} \\
\multicolumn{1}{c|}{}                                  & \textbf{Synthesized} & 2                                   & 2             & 84.292                               &                        \\ \hline \hline
\end{tabular}
\label{tab:all_basic_class}
\vspace{-10pt}
\end{table}

\subsection{Error Modes}

We recognize that the majority of errors encountered in quantum computing can be effectively modeled through bit-flip and phase-flip errors. Bit-flip errors alter the state of a qubit from $\ket{0}$ to $\ket{1}$ or vice versa, akin to flipping a bit in classical computing. Phase-flip errors, on the other hand, affect the phase of the qubit, which is a quantum property without a classical counterpart.

In this research, we focus on three primary error modes: depolarizing errors, a blend of bit-flip and phase-flip errors, and a comprehensive model that combines all three error types. Depolarizing errors represent a more generalized form of quantum noise, where a qubit state is randomized, potentially leading to the loss of its original information. Each of these error models is associated with a specific probability, indicating the likelihood of the error being applied across the entire circuit. This probabilistic approach allows us to simulate the impact of quantum errors on our systems with a realistic variance, providing insights into how these errors can affect quantum computing operations and the effectiveness of our error correction strategies.

\subsection{QECC Implementation}

\begin{table}[]
\centering
\caption{Impact of QECC implementation on circuit characteristics.}
\begin{tabular}{ccc||ccc}
\multicolumn{3}{c||}{\textbf{Complete Circuit}}                                                                             & \multicolumn{3}{c}{\textbf{Properties}}                                                \\ \hline \hline
\multicolumn{1}{c|}{\textbf{Classifier}}                & \multicolumn{1}{c|}{\textbf{Class}}              & \textbf{QECC} & \multicolumn{1}{c|}{\textbf{Qubits}} & \multicolumn{1}{c|}{\textbf{Gates}} & \textbf{Depth} \\ \hline
\multicolumn{1}{c|}{\multirow{8}{*}{\textbf{1-qubit}}}  & \multicolumn{1}{c|}{\multirow{4}{*}{\textbf{M}}} & None          & 1                                    & 3                                   & 3              \\
\multicolumn{1}{c|}{}                                   & \multicolumn{1}{c|}{}                            & Steane        & 10                                   & 119                                 & 53             \\
\multicolumn{1}{c|}{}                                   & \multicolumn{1}{c|}{}                            & D3Surface     & 17                                   & 149                                 & 44             \\
\multicolumn{1}{c|}{}                                   & \multicolumn{1}{c|}{}                            & D5Surface     & 36                                   & 167                                 & 59             \\ \cline{2-6} 
\multicolumn{1}{c|}{}                                   & \multicolumn{1}{c|}{\multirow{4}{*}{\textbf{C}}} & None          & 1                                    & 4                                   & 4              \\
\multicolumn{1}{c|}{}                                   & \multicolumn{1}{c|}{}                            & Steane        & 10                                   & 126                                 & 54             \\
\multicolumn{1}{c|}{}                                   & \multicolumn{1}{c|}{}                            & D3Surface     & 17                                   & 152                                 & 45             \\
\multicolumn{1}{c|}{}                                   & \multicolumn{1}{c|}{}                            & D5Surface     & 36                                   & 170                                 & 59             \\ \hline \hline
\multicolumn{1}{c|}{\multirow{16}{*}{\textbf{2-qubit}}} & \multicolumn{1}{c|}{\multirow{4}{*}{\textbf{R}}} & None          & 2                                    & 7                                   & 3              \\
\multicolumn{1}{c|}{}                                   & \multicolumn{1}{c|}{}                            & Steane        & 17                                   & 241                                 & 81             \\
\multicolumn{1}{c|}{}                                   & \multicolumn{1}{c|}{}                            & D3Surface     & 26                                   & 306                                 & 53             \\
\multicolumn{1}{c|}{}                                   & \multicolumn{1}{c|}{}                            & D5Surface     & 72                                   & 352                                 & 76             \\ \cline{2-6} 
\multicolumn{1}{c|}{}                                   & \multicolumn{1}{c|}{\multirow{4}{*}{\textbf{B}}} & None          & 2                                    & 8                                   & 4              \\
\multicolumn{1}{c|}{}                                   & \multicolumn{1}{c|}{}                            & Steane        & 17                                   & 248                                 & 81             \\
\multicolumn{1}{c|}{}                                   & \multicolumn{1}{c|}{}                            & D3Surface     & 26                                   & 309                                 & 53             \\
\multicolumn{1}{c|}{}                                   & \multicolumn{1}{c|}{}                            & D5Surface     & 72                                   & 355                                 & 76             \\ \cline{2-6} 
\multicolumn{1}{c|}{}                                   & \multicolumn{1}{c|}{\multirow{4}{*}{\textbf{G}}} & None          & 2                                    & 8                                   & 4              \\
\multicolumn{1}{c|}{}                                   & \multicolumn{1}{c|}{}                            & Steane        & 17                                   & 248                                 & 81             \\
\multicolumn{1}{c|}{}                                   & \multicolumn{1}{c|}{}                            & D3Surface     & 26                                   & 309                                 & 53             \\
\multicolumn{1}{c|}{}                                   & \multicolumn{1}{c|}{}                            & D5Surface     & 72                                   & 355                                 & 76             \\ \cline{2-6} 
\multicolumn{1}{c|}{}                                   & \multicolumn{1}{c|}{\multirow{4}{*}{\textbf{Y}}} & None          & 2                                    & 9                                   & 4              \\
\multicolumn{1}{c|}{}                                   & \multicolumn{1}{c|}{}                            & Steane        & 17                                   & 255                                 & 81             \\
\multicolumn{1}{c|}{}                                   & \multicolumn{1}{c|}{}                            & D3Surface     & 26                                   & 312                                 & 53             \\
\multicolumn{1}{c|}{}                                   & \multicolumn{1}{c|}{}                            & D5Surface     & 72                                   & 358                                 & 76             \\ \hline \hline
\end{tabular}
\label{tab:all_circuit_table}
\end{table}

In our study, we focus on three specific QECCs: the Steane code \cite{steane1996multiple} and two variations of the surface code, characterized by distances of $3$ and $5$ \cite{krinner2022realizing}. On one hand, the selection of the Steane code is crucial due to its unique ability to correct both bit-flip and phase-flip errors simultaneously with a relatively simple lattice structure, making it an ideal candidate for demonstrating fault-tolerant quantum computation. On the other hand, the surface code is recognized as the most feasible QECC currently available, boasting a notably high error threshold. This high threshold makes the surface code particularly attractive for practical quantum computing applications, as it suggests a greater tolerance for errors before the integrity of quantum information is compromised. We applied QECCs on the classifiers using the `MQT-QECC' framework from the Munich Quantum Toolkit \cite{grurl2023automatic} and simulated the integrated circuits with IBM Qiskit's `AerSimulator'.

\begin{figure}
    \centering
    \includegraphics[width=1\linewidth]{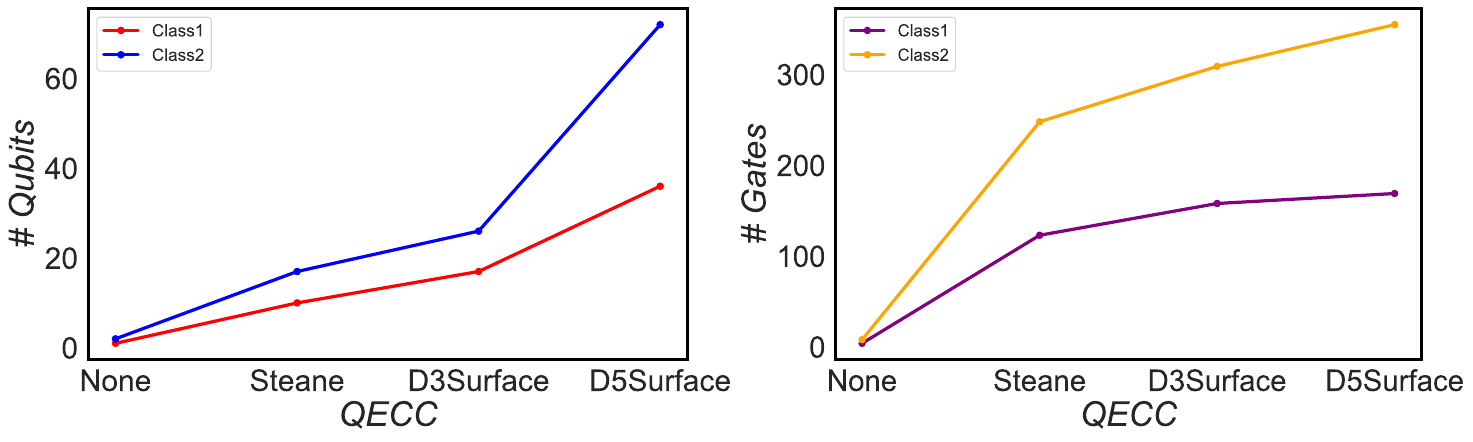}
    \vspace{-20pt}
    \caption{\textbf{Overhead analysis post-QECC application:} 
    This figure illustrates the average increase in qubit count (left) and gate count (right) required for the classifiers after the integration of QECCs.
    }
    \label{fig:overhead}
    \vspace{-10pt}
\end{figure}

As previously mentioned, the one-qubit classifier is tasked with distinguishing between two color categories (`M' \& `C') within a two-dimensional dataset. In contrast, the two-qubit classifier is designed to differentiate among four color categories (`R', `B', `G' \& `Y') in a four-dimensional dataset. We select a representative `point' from each category and subject it to QECCs. Each point, requiring classification, is processed through a unique amplitude encoding circuit. We incorporate QECCs within a composite circuit encompassing both the classifier and amplitude encoding circuits. This integration is critical for assessing the additional complexity introduced by implementing QECCs, referred to as the overhead. Table \ref{tab:all_circuit_table} outlines the characteristics of the circuits post-application of QECCs for each point across the classes, providing insight into the implications of QECC deployment on circuit properties. Fig. \ref{fig:overhead} shows the average qubit (left) and gate (right) overhead of the classifiers post-QECC application.

\subsection{Evaluation Metrics}

We evaluate QECC-enhanced quantum classifiers through a dual-phase analysis, initially examining chosen reference points per class. For the one-qubit classifier, we use two reference points (one per class) to build a circuit with amplitude encoding, the classifier, and QECC, aiming for binary measurement outcomes ($0$ or $1$). For the two-qubit classifier, four reference points (one per class) are used, targeting measurement outcomes of $00$, $01$, $10$, and $11$ for each class. We test both classifiers' resilience by sweeping the noise levels, error modes, and QECC types, focusing on the probability of successful trials (PST) for accurate classification of each selected point.

Next, we present a theoretical analysis of the impact of noise on the accuracy of a classifier within a synthetic dataset where each class is represented by an equal number of data points. We use a controlled environment where noise affects all classes uniformly. A classifier's purpose is to correctly identify the class to which a data point belongs. In an ideal scenario without noise, we denote the probability of correctly classifying a data point in class \( c_i \) as \( p_{ci} \). However, real-world scenarios are rarely ideal, and classifiers must contend with noise that can degrade performance. We denote the probability of correctly classifying a data point in the presence of noise as \( p'_{ci} \) and the decrease in classification probability due to noise as \( \Delta p_{ci} = p_{ci} - p'_{ci} \). Under the assumption of uniform noise distribution, the impact of noise on the classification accuracy for each class is the same, and thus the overall accuracy of the classifier can be expressed as:
\[
A' = \frac{1}{n} \sum_{i=1}^{n} p'_{ci} = \frac{1}{n} \sum_{i=1}^{n} (p_{ci} - \Delta p) = A - \Delta p
\]
where \( A \) is the original accuracy of the classifier without noise, \( A' \) is the accuracy of the classifier with noise, \( n \) is the number of classes, and \( \Delta p \) is the uniform decrease in classification probability due to noise. 
This model shows that minor, consistent drops in classifying individual points lower overall classifier accuracy proportionally. Thus, if noise \( \gamma \) reduces classification success by $\Delta p\%$ uniformly across all classes, the classifier's total accuracy is expected to fall by $\Delta p\%$, reflecting the impact of noise \( \gamma \).
\section{Experiments and Results} \label{sec:exp_result}

\begin{figure}
    \centering
    \includegraphics[width=1\linewidth]{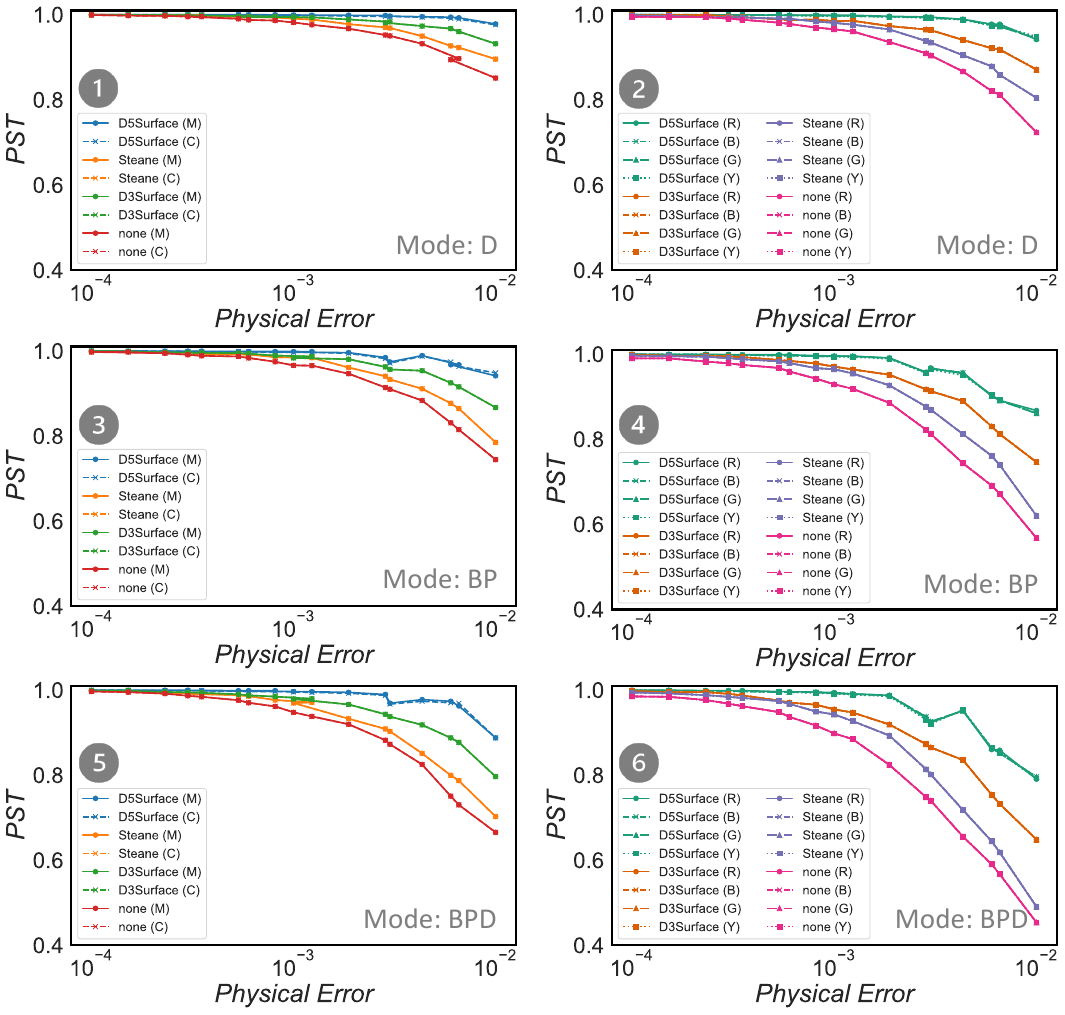}
    \vspace{-20pt}
    \caption{\textbf{Comparative performance of QECCs under physical noise:} 
     This figure displays QECC performance for 1-qubit and 2-qubit classifiers under `D', `BP', and `BPD' error modes as physical noise increases. With six subfigures, it compares resilience between classifiers, highlighting the distance $5$ surface code's superior improvement and the Steane code's minimal gain across noise levels.
    }
    \label{fig:pst_with_all_points}
\end{figure}

\subsection{Impact of Physical Errors}

Our initial experiments explore QECC responses to various physical error rates, capping at the surface codes' correction threshold of $10^{-2}$. Beyond this threshold, data becomes less relevant, guiding our limit on error rate variations. We initially chose reference points `M' and `C' for the first classifier and `R', `B', `G', and `Y' for the second. Our focus now shifts to how their success probabilities vary with increased physical errors, using different QECCs and error modes. Results in Fig. \ref{fig:pst_with_all_points} highlight QECC resilience and efficiency against rising error rates.
\begin{figure}
    \centering
    \includegraphics[width=1\linewidth]{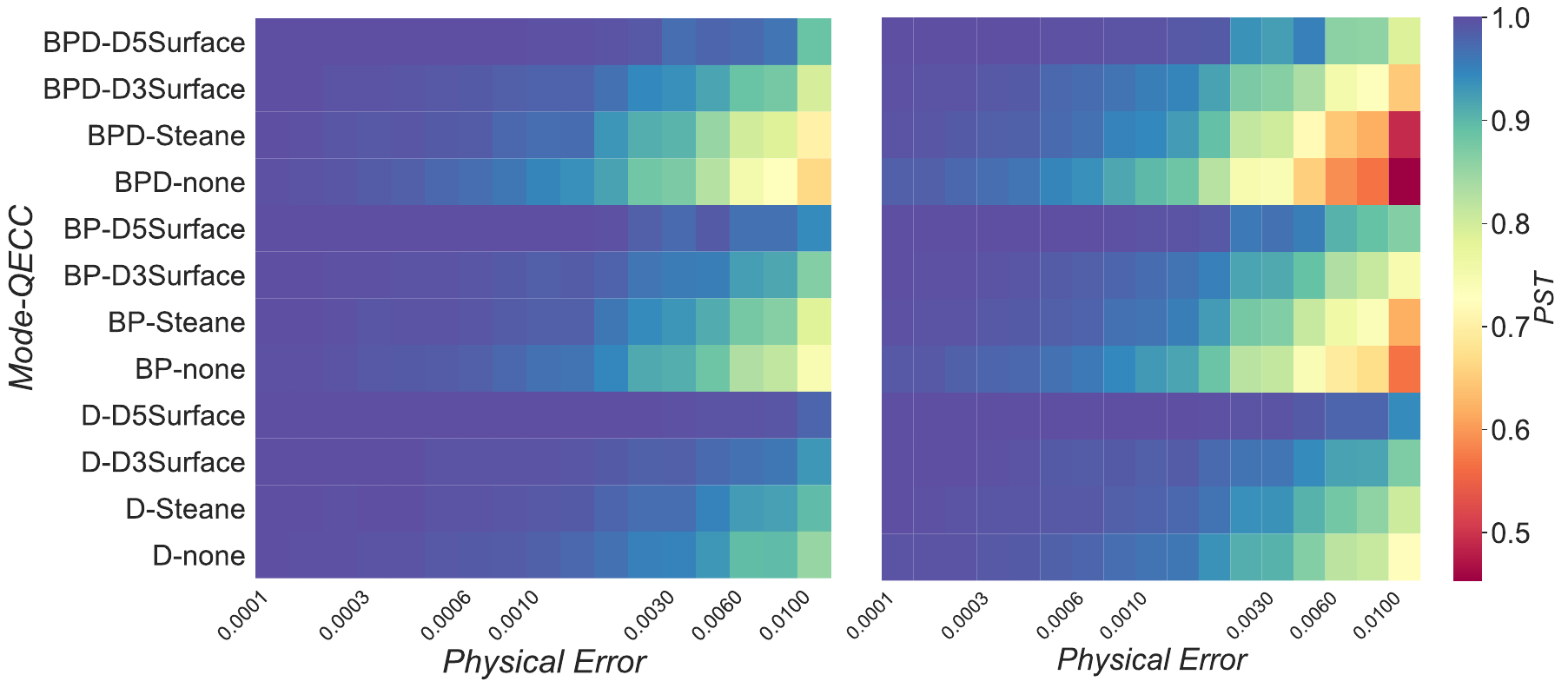}
    \vspace{-20pt}
    \caption{\textbf{Heatmap analysis of quantum classifier resilience:} 
    This figure presents a heatmap comparison of 1-qubit (left) and 2-qubit (right) classifiers, showing how physical noise affects their performance. The 2-qubit classifier shows a sharper drop in successful trial probability (PST) with higher noise levels, particularly under BPD and least with D.
    }
    \label{fig:heatmap_pst}
    \vspace{-10pt}
\end{figure}
Fig. \ref{fig:pst_with_all_points} is organized into two columns for the two classifiers and three rows for error modes `D', `BP', and `BPD', resulting in six subfigures. These subfigures chart the success probabilities of reference plots against increasing noise levels. For the 1-qubit classifier (subfigures
\raisebox{.5pt}{\textcircled{\raisebox{-.9pt} {\textbf{1}}}}, 
\raisebox{.5pt}{\textcircled{\raisebox{-.9pt} {\textbf{3}}}} \& 
\raisebox{.5pt}{\textcircled{\raisebox{-.9pt} {\textbf{5}}}}),
the distance $5$ surface code shows the most significant improvement, theoretically expected to outperform distance $3$ and then the Steane code, which shows the least improvement. The 2-qubit classifier (subfigures 
\raisebox{.5pt}{\textcircled{\raisebox{-.9pt} {\textbf{2}}}}, 
\raisebox{.5pt}{\textcircled{\raisebox{-.9pt} {\textbf{4}}}} \& 
\raisebox{.5pt}{\textcircled{\raisebox{-.9pt} {\textbf{6}}}}
) exhibits a similar trend.
The second classifier, using more qubits and gates, experiences greater error propagation, resulting in lower success probabilities compared to the first. Despite this, an in-depth analysis shows similar PST patterns for both classifiers. Thus, we focus on a specific reference point for each going forward: `M' for the first and `R' for the second.

Fig. \ref{fig:heatmap_pst} uses heatmaps to compare 1-qubit (left) and 2-qubit (right) classifiers, showing a decline in PST with more physical noise. The 2-qubit classifier's PST drops more due to higher error propagation. Of the error modes, the combination of bit-, phase-flip, and depolarizing errors affect PST most, followed by the duality of bit and phase-flip errors, with depolarizing errors alone impacting the least.

The error modes `D' (Depolarizing), `BP' (mix of Bit-flip and Phase-flip), and `BPD' (mix of Bit-flip, Phase-flip, and Depolarizing) uniquely affect quantum system performance. BPD is notably the most harmful, causing comprehensive qubit state corruption, including state switches, phase alterations, and information erasure. This combination makes BPD the toughest challenge for error correction, resulting in significant performance degradation.
BP mode distorts computational basis states and superpositions without causing full state randomization like depolarizing errors. Consequently, quantum information is corrupted but not entirely lost or turned into a mixed state, making BP's impact substantial yet less severe than BPD mode.
D mode randomizes qubit states, causing quantum information loss. Yet, its effects are predictable, making correction simpler than for the mixed errors in BP and BPD modes. Depolarizing errors broadly affect the qubit state, making their impact less intricate than the combined effects in BP and BPD. Thus, BPD is the most harmful, followed by BP, while D is easier to manage for error correction.

\subsection{Classifier Performance with various QECCs}

\begin{figure}
    \centering
    \includegraphics[width=1\linewidth]{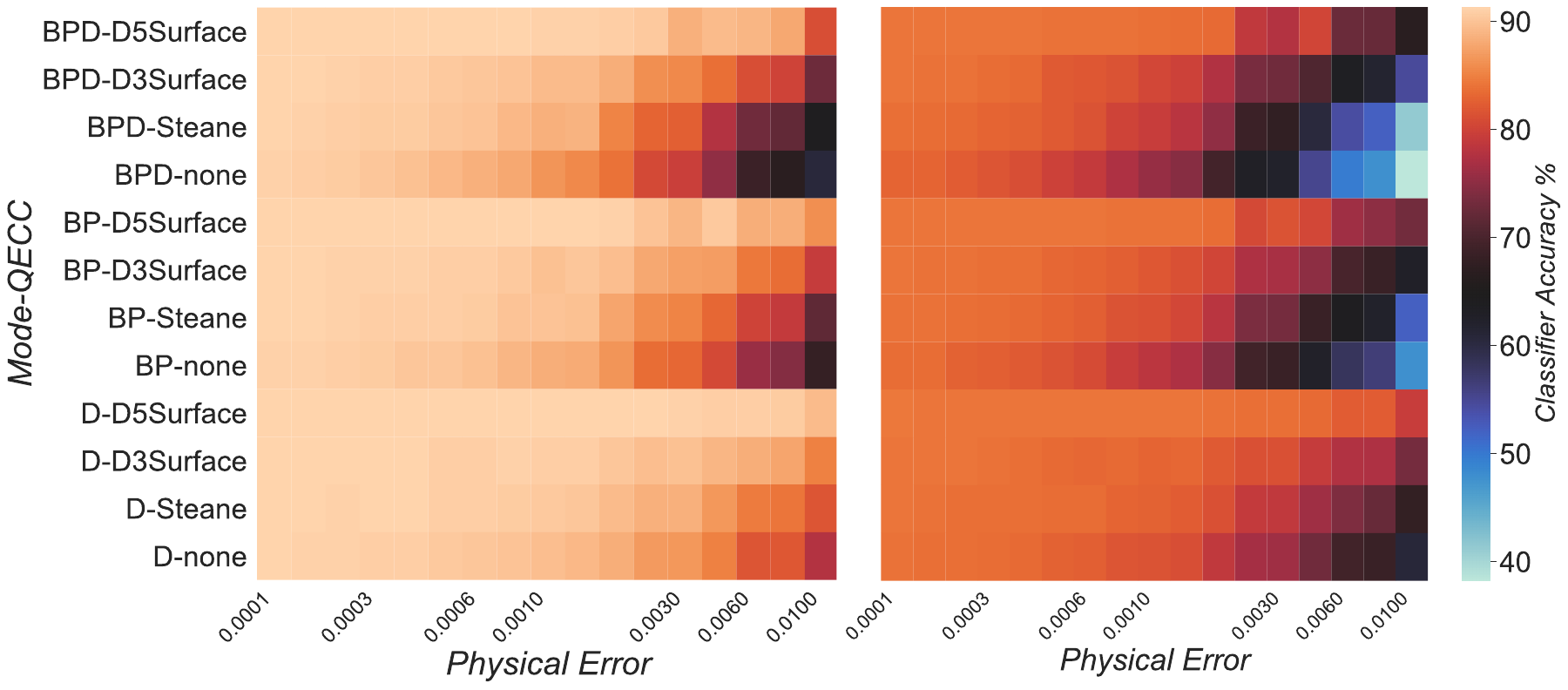}
    \vspace{-20pt}
    \caption{\textbf{Heatmap analysis of quantum classifier performance:} 
    This figure illustrates a side-by-side heatmap analysis contrasting the accuracy of the 1-qubit and 2-qubit classifiers. This visual representation aids in understanding the differential impact of various factors on each classifier's performance.
    }
    \label{fig:heatmap_accuracy}
    \vspace{-10pt}
\end{figure}

After examining quantum classifiers' performance across different noise conditions, we now turn to how QECCs help sustain accuracy. Originally, the classifiers had accuracies of $91.33\%$ and $84.23\%$. It is important to recognize that QECCs do not directly improve performance but mitigate noise effects on accuracy. 

Our analysis begins with a comparative heatmap in Fig. \ref{fig:heatmap_accuracy}, showcasing the 1-qubit classifier on the left and the 2-qubit classifier on the right. Generally, the 2-qubit classifier exhibits lower accuracy compared to the 1-qubit classifier, a trend consistent with their performances under ideal conditions. Upon examining each noise mode, we observe a notable enhancement in accuracy with the application of QECCs as opposed to scenarios where no QECC is utilized. As anticipated, the distance $5$ surface code demonstrates the greatest improvement, whereas the Steane code offers the least. Consistent with previous discussions, the `BPD' noise model proves to be the most harmful across all scenarios, with `D' being the least impactful on accuracy.

\begin{figure}
    \centering
    \includegraphics[width=1\linewidth]{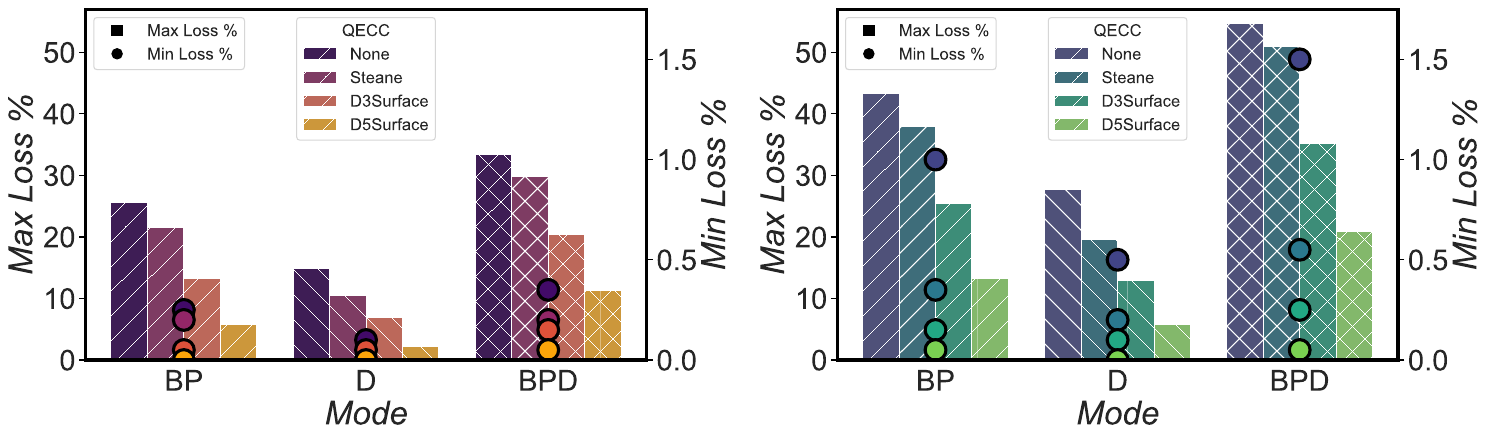}
    \vspace{-20pt}
    \caption{\textbf{Classifier accuracy loss after QECC addition:} 
    This figure shows accuracy losses for 1-qubit (left) and 2-qubit (right) classifiers using QECCs, revealing how different error modes, especially `BPD', affect performance. 
    }
    \label{fig:compare_loss}
    \vspace{-10pt}
\end{figure}

Having examined how accuracy varies between different classifiers and QECCs, we now turn our attention to quantifying the benefits QECCs offer in terms of accuracy. Fig. \ref{fig:compare_loss} presents a comparison of the minimum and maximum accuracy loss experienced by classifiers upon integrating QECCs, for both the 1-qubit (left) and 2-qubit (right) classifiers. The 2-qubit classifier shows a greater accuracy loss than the 1-qubit classifier, attributed to its increased complexity, and the higher number of qubits and gates involved. Among the error modes, `BPD' proves to be the most damaging, leading to the largest accuracy loss, whereas `D' results in the least. Regarding QECCs, the distance $5$ surface code outperforms others by exhibiting the smallest loss, in contrast to the Steane code, which incurs the highest loss beyond the point where no QECCs are applied. Table \ref{tab:average_improvement} provides a summary of classifier performance enhancements following QECC implementation. It details the average accuracy improvements for each classifier across various physical noise levels, noise modes, and QECC types. Additionally, it quantifies the percentage increase in accuracy each QECC offers over classifiers without QECC implementation. Consistent with expectations, the distance $5$ surface codes yield the most significant improvement, while the Steane code results in the least.

\begin{table}[]
\centering
\caption{Classifier enhancements post-QECC implementation: Average Accuracy (AA) and Accuracy Improvement (AI)}
\begin{tabular}{c|c||cc||cc}
\multirow{2}{*}{\textbf{Mode}} & \multirow{2}{*}{\textbf{QECC}} & \multicolumn{2}{c||}{\textbf{1-qubit Class.}}         & \multicolumn{2}{c}{\textbf{2-qubit Class.}}          \\ \cline{3-6} 
                               &                                & \multicolumn{1}{c|}{\textbf{AA \%}} & \textbf{AI \%} & \multicolumn{1}{c|}{\textbf{AA \%}} & \textbf{AI \%} \\ \hline \hline
\multirow{4}{*}{\textbf{D}}    & \textbf{None}                  & 87.82                               & NA             & 78.28                               & NA             \\
                               & \textbf{Steane}                & 88.98                               & 1.32           & 80.23                               & 2.49           \\
                               & \textbf{D3Surface}             & 89.99                               & 2.47           & 81.68                               & 4.34           \\
                               & \textbf{D5Surface}             & 91.05                               & 3.68           & 83.60                               & 6.80           \\ \hline
\multirow{4}{*}{\textbf{BP}}   & \textbf{None}                  & 85.36                               & NA             & 73.51                               & NA             \\
                               & \textbf{Steane}                & 86.98                               & 1.90           & 76.44                               & 3.99           \\
                               & \textbf{D3Surface}             & 88.62                               & 3.82           & 78.93                               & 7.37           \\
                               & \textbf{D5Surface}             & 90.31                               & 5.80           & 81.87                               & 11.37          \\ \hline
\multirow{4}{*}{\textbf{BPD}}  & \textbf{None}                  & 82.69                               & NA             & 69.61                               & NA             \\
                               & \textbf{Steane}                & 84.59                               & 2.30           & 72.79                               & 4.57           \\
                               & \textbf{D3Surface}             & 87.17                               & 5.42           & 76.39                               & 9.74           \\
                               & \textbf{D5Surface}             & 89.92                               & 8.74           & 80.69                               & 15.92          \\ \hline \hline
\end{tabular}
\label{tab:average_improvement}
\end{table}

\subsection{Practical Implications}

\begin{figure}
    \centering
    \includegraphics[width=1\linewidth]{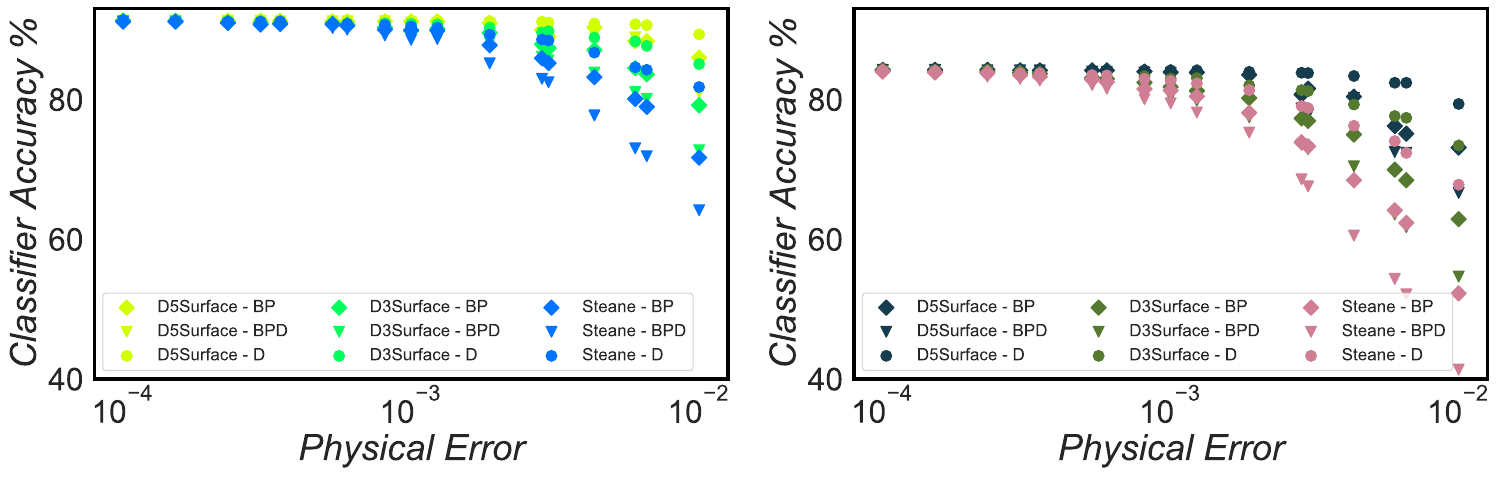}
    \vspace{-20pt}
    \caption{\textbf{Strategic QECC selection for optimized classifier performance:} 
    This figure contrasts one-qubit (left) and two-qubit (right) classifier accuracies under different QECCs and error modes. It shows occasions where the Steane code surpasses the surface codes .
    }
    \label{fig:practical_accuracy_scatter}
    \vspace{-10pt}
\end{figure}

Our observations have shown that QECCs can significantly enhance the performance of quantum classifiers in noisy environments. Among the codes evaluated, the distance $5$ surface code offers the most substantial improvement, followed by the distance $3$ surface code and then the Steane code. To delve deeper, we further analyze the accuracies of both classifiers across different QECCs and error modes, this time presenting the data in scatter plots. Figure \ref{fig:practical_accuracy_scatter} displays the accuracy comparisons for the one-qubit classifier (left) and the two-qubit classifier (right).

A careful examination of the scatter plots reveals instances where the Steane code outperforms the distance $3$ surface code at certain noise levels. This variation suggests that the advantage of a specific QECC depends on the nature of error modes present at particular levels of physical noise. Specifically, for the two-qubit classifier at a physical noise level of $10^{-2}$, the Steane code, when dealing with error mode `D', surpasses the distance $3$ surface code in mode `BP' and even edges out the distance $5$ surface code in mode `BPD'. This indicates that in practical scenarios, where minimizing the number of qubits and gates is crucial, understanding the specific type and level of noise could enable more efficient QECC selection, potentially reducing overhead while preserving accuracy. Hence, the optimal choice of QECC is not necessarily the most robust code under all conditions but rather the code best suited to the particular circumstances encountered.

\section{Limitations and Challenges} \label{sec:lim}

This study deliberately limits its focus to employing only one and two-qubit classifiers because increasing the number of qubits in the base circuit significantly escalates the number of qubits required in the QECC circuits, which in turn affects simulation times adversely. While incorporating additional gates into the classifiers can improve their performance, this enhancement comes with similar challenges as observed with QECCs: a substantial increase in the number of gates and circuit depth. Moreover, integrating multi-qubit gates within a circuit poses a significant challenge for most QECCs, necessitating the use of complex transversal gates like lattice surgery, which are notably intricate to implement. In practical scenarios, classifiers will encounter all these limitations, making the application of QECCs to them a complex endeavor. Nevertheless, this research provides valuable insights into what can be anticipated and represents a pioneering effort, laying the groundwork for future explorations.
\section{Conclusion} \label{sec:conclusion}

This research applies QECCs to enhance quantum classifiers for the first time. Using two synthetic datasets namely, a two-dimensional and a four-dimensional, we developed and tested one-qubit and two-qubit classifiers, respectively, against three error modes: depolarizing, bit and phase flips, and a combined model. We explored three QECCs: Steane code and surface codes at distances of $3$ and $5$, finding the distance $5$ surface codes most effective in error correction. The `BPD' error mode was identified as the most harmful. Our research indicates that while theoretical assessments can highlight one QECC as superior to others, the choice of an optimal QECC in real-world applications hinges on the specific context, including constraints on qubit availability, required accuracy levels, and the nature and intensity of physical errors. This insight lays the groundwork for future explorations and applications of QECCs in quantum computing, emphasizing the importance of a nuanced approach to selecting QECCs based on the unique demands of each quantum computing task.



\bibliographystyle{unsrt}
\bibliography{refs}


\end{document}